\documentclass[journal=jacsat,manuscript=article]{achemso}

\usepackage[version=3]{mhchem} 
\usepackage{textcomp,mathcomp}

\author{Qi Ou}
\altaffiliation{Contributes equally to this work}
\affiliation{SINOPEC Research Institute of Petroleum Processing Co., Ltd, Beijing 100083, China.}
\email{ouqi.ripp@sinopec.com}
\author{Hongshuai Wang}
\altaffiliation{Contributes equally to this work}
\affiliation{DP Technology, Beijing 100080, China}
\author{Minyang Zhuang}
\affiliation{SINOPEC Research Institute of Petroleum Processing Co., Ltd, Beijing 100083, China.}
\author{Shangqian Chen}
\affiliation{DP Technology, Beijing 100080, China}
\author{Lele Liu}
\author{Ning Wang}
\email{wangning.ripp@sinopec.com}
\affiliation{SINOPEC Research Institute of Petroleum Processing Co., Ltd, Beijing 100083, China.}
\author{Zhifeng Gao}
\email{gaozf@dp.tech}
\affiliation{DP Technology, Beijing 100080, China}
\alsoaffiliation{AI for Science Institute, Beijing 100084, China}

\title[An \textsf{achemso} demo]
  {High-Accuracy Physical Property Prediction for Organics via Molecular Representation Learning: Bridging Data to Discovery}

\begin{document}

\begin{abstract}
The ongoing energy crisis has underscored the urgent need for energy-efficient materials with high energy utilization efficiency, prompting a surge in research into organic compounds due to their environmental compatibility, cost-effective processing, and versatile modifiability. To address the high experimental costs and time-consuming nature of traditional trial-and-error methods in the discovery of highly functional organic compounds, we apply the 3D transformer-based molecular representation learning algorithm to construct a pre-trained model using 60 million semi-empirically optimized structures of small organic molecules, namely, Org-Mol, which is then fine-tuned with public experimental data to obtain prediction models for various physical properties. Despite the pre-training process relying solely on single molecular coordinates, the fine-tuned models achieves high accuracy (with $R^2$ values for the test set exceeding 0.95). These fine-tuned models are applied in a high-throughput screening process to identify novel immersion coolants among millions of automatically constructed ester molecules, resulting in the experimental validation of two promising candidates. This work not only demonstrates the potential of Org-Mol in predicting bulk properties for organic compounds but also paves the way for the rational and efficient development of ideal candidates for energy-saving materials.
\end{abstract}

\section{Introduction}
Amidst the current energy crisis, the research and development of materials that are energy-efficient and/or offer high energy utilization efficiency have become an immediate imperative and a focal point in scientific inquiry.\cite{science.adq3799,davidson2019exnovating,zhang2014organic,Kandpal2024} Organic compounds play a pivotal role in this context, owing to their environmental friendliness, low-cost processing, and almost infinite modifiability to meet a wide array of property requirements. Prospective applications lie in the fields such as immersion coolants for data centers,\cite{kuncoro2019immersion,kanbur2021system,huang2024experimental} phase change materials for thermal energy storage,\cite{sharma2015developments,li2019colossal} and liquid organic hydrogen carriers.\cite{preuster2017liquid,mah2021targeting,tan2024advances} All of these fields raise specific requirements on various physical and chemical properties of the applied organics. Taking the immersion coolants for the data center as an example, an ideal coolants should possess low dielectric constant and viscosity as well as high heat capacity and thermal conductivity, so as to provide a high thermal exchange efficiency during the circulation while prevent plausible signal loss during the operation of electronic devices.\cite{chhetri2022numerical,wu2023prediction} 


While the measurement of the aforementioned properties 
for a single compound is straightforward,\cite{nadkarni2007guide} the synthesis 
of various candidates can be time-consuming. Therefore, the experimental trial-and-error cost of exploring the chemical space in order to develop ideal energy-saving organic materials remains high. Multiple attempts have been made to predict these electrical, mechanical, and thermal properties of organic compounds via molecular dynamics (MD) and/or machine learning (ML) methods.\cite{yamaguchi2020simulations,caleman2012force,ju2016prediction,sild2002general,meng2014neural,deng2022dielectric,luo2024physical,zheng2023molecular,shan2024develop} However, the accuracy of such methods relies on the applied force field and/or the algorithm applied in the ML model as well as the quality and the size of the data set.\cite{gereben2011accurate, krishnamoorthy2021dielectric,comesana2022systematic} It can be seen that predicting these physical properties via MD is time-consuming for a single system\cite{yamaguchi2020simulations} and become unaffordable in a high-throughput screening scenario with thousands of candidates, while the validity of some previously reported ML models is limited to a few specific types of compounds.\cite{PhysRevB.110.165159,zheng2023molecular,meng2014neural} The requirement of constructing the appropriate descriptors for ML models can also be computationally expensive and sometimes non-trivial,\cite{sild2002general,deng2022dielectric,comesana2022systematic,gao2019artificial}  raising additional challenges for an accurate yet efficient prediction which is generalizable to a plethora of organic compounds.

With the development of deep learning algorithms, highly efficient screening and rational design of novel material with required properties have been further made simple.\cite{walters2020applications,meftahi2020machine,pang2023advanced} For instance, Uni-Mol, a 3D transformer-based molecular representation learning (MRL) algorithm,\cite{zhou2023unimol} has been successfully applied in the prediction of photophysical properties of organic photoluminescence systems\cite{cheng2023automatic} and the adsorption energies of various gaseous molecules on metal-organic-frameworks.\cite{wang2024comprehensive} One of the most favored advantages of Uni-Mol is that sufficient and rational descriptors can be automatically constructed via the pre-training process, requiring only the single molecular structures and a small number of labeled data as input.\cite{zhou2023unimol,cheng2023automatic,wang2024comprehensive} While these previously reported properties predicted by the Uni-Mol framework depend solely on a single molecular structure or on a single cell of a periodic crystallized structure, its feasibility in predicting bulk properties for amorphous (or liquid phase) organic compounds is yet to be validated. If proven effective, such protocol may remarkably streamline the model-constructing process and reduce the computational cost in predicting a variety of physical properties for amorphous organics.  

Keeping these in mind and to facilitate the discovery of energy-saving materials in terms of organic liquids, we apply the Uni-Mol framework to develop a pre-trained model specifically designed for organic compounds, namely, the Org-Mol model, followed by the fine-tuning process over varied physical properties with public experimental data. Our results evince that, even though only the single molecular coordinates are provided during the pre-training process, the fine-tuned models for all tested bulk properties are of significantly high accuracy (with the $R^2$ value of the test set higher than 0.95). Furthermore, these fine-tuned models are applied in a high-throughput screening for the discovery of novel immersion coolants over millions of ester molecules. Hundreds of candidates are finally filtered out and two of them are experimentally synthesized and validated, highlighting the capability of Org-Mol to directly bridge data to discovery. While the prediction model for other physical properties can be straightforwardly established via the same fine-tuning process, the ideal candidates for not only the immersion coolants but other energy saving materials can be therefore rationally and efficiently developed.  

\section{Computational Details}
\begin{figure}[h]
  \centering
  \includegraphics[width=16.5cm]{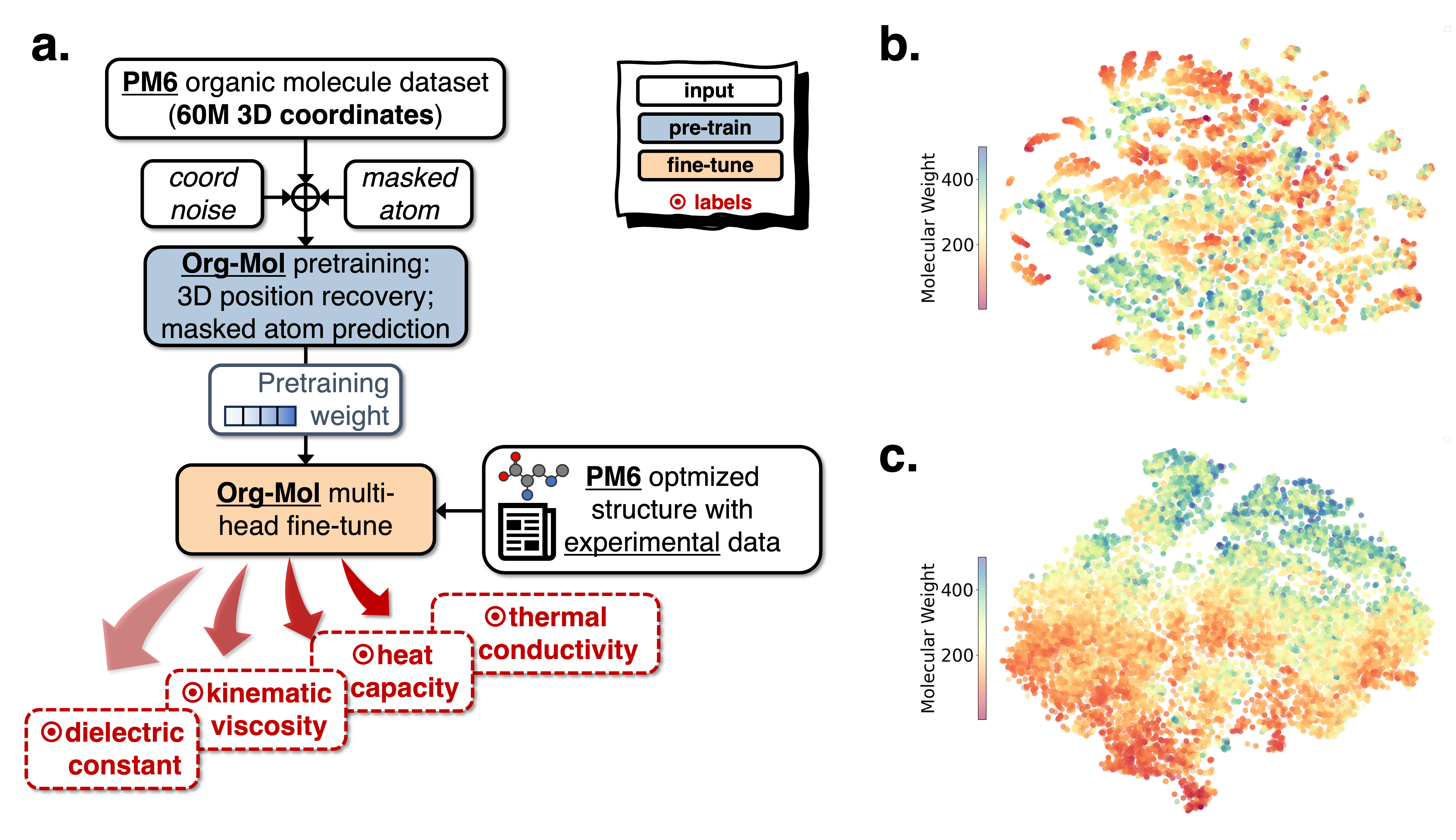}
  \caption{$\bf a.$ Schematic graph of the pre-training and fine-tuning process of the Org-Mol model for organic compounds. $\bf b$, $\bf c.$ Distributions of the molecular weight of the pre-training dataset in 2D projections of Org-Mol embedding layer via t-SNE. Data points in $\bf b$ are obtained with a randomly initialized weight while those in $\bf c$ are obtained with the Org-Mol pretrained weight.}
  \label{fig:schema}
\end{figure}

We apply the framework of Uni-Mol to construct a pre-trained model particularly designed for organic molecules containing C, H, O, N, S, Se, B, and halogens, namely, the Org-Mol pre-trained model. This pre-trained model is constructed based on the PubChemQC PM6 data sets.\cite{Nakata2019PubChemQCPD} (See Ref.~\citenum{zhou2023unimol} for the detailed architecture of the pre-training process.) Specifically, PM6\cite{stewart2007optimization}-optimized structures of 60 million small organic molecules containing the aforementioned elements are fed into thepre-training framework to initiate the self-supervised task including the 3D coordinates recovery and masked atom prediction. The Org-Mol pre-trained model is then fine-tuned via multiple experimentally measured physical properties of general organic compounds, i.e., the dielectric constant near 25 \textcelsius\ collected from Ref.~\citenum{haynes2014crc} and Ref.~\citenum{LandoltBornstein1991:sm_lbs_978-3-540-47619-1_2}, the kinematic viscosity near 40 and 100 \textcelsius\ collected from Ref.~\citenum{yaws1999chemical} and Ref.~\citenum{LandoltBornstein2017:sm_lbs_978-3-662-49218-5_364}, the density at multiple temperatures and two thermal properties at 25 \textcelsius\ (thermal conductivity and heat capacity) collected from Ref.~\citenum{yaws1999chemical}. These data together with the PM6-optimized single molecular structures of the corresponding compounds are collected and divided into training set and test set in a 9:1 ratio. The whole model-construction process is schematized in Figure \ref{fig:schema}$\bf a$. Metrics applied to validate the fine-tuned model for the aforementioned properties include the $R^2$ value, the mean absolute error (MAE), the root mean square error (RMSE), and the mean absolute percentage error (MAPE) are calculated as follows
\begin{eqnarray}
R^2 &=& 1 - \frac{\sum^N_{i=1}(\hat y_i-y_i)^2}{\sum^N_{i=1}(\hat y_i-\bar y)^2} \\
\textrm{MAE} &=& \frac{1}{N} \sum^N_{i=1} |\hat y_i-y_i| \\
\textrm{MAPE(\%)} &=&  \frac{1}{N} \sum^N_{i=1} \frac {|\hat y_i-y_i|}{\textrm{max}(\eta,|y_i|)} \times 100\% \\
\textrm{RMSE} &=& \sqrt{\frac{1}{N} \sum^N_{i=1} (\hat y_i-y_i)^2}
\end{eqnarray}
where $N$ is the number of samples, $y_i$ the experimentally measured data for sample $i$, $\hat y_i$ the Org-Mol predicted data for sample $i$, $\bar y$ the average of the experimental data, $\eta$ an arbitrary small yet strictly positive number to avoid undefined results when $y_i$ is zero.

To investigate the relationship between the single molecular dipole moment and the dielectric constant of the corresponding liquid phase system, we calculate the dipole moment for randomly picked systems in the dielectric constant data set via density functional theory (DFT) in quantum chemistry package Gaussian 16.\cite{g16} B3LYP functional (with D3 empirical dispersion correction) and def2-svp basis set are applied to perform the single-molecule ground state geometry optimization and dipole moment calculation. 

\section{Results and discussion}
We first perform a t-SNE analysis\cite{van2008visualizing} for the molecular weight of the pre-training data set to speculate the effectiveness of Org-Mol pre-trained model in learning molecular structures and its ability to distinguish between different molecules. The 512-dimension embedding layer
of the Org-Mol model is projected into two dimensions, and the distributions of the molecular weight of 20,000 randomly selected molecules in the pre-training data are shown in Figure \ref{fig:schema}$\bf b$ and $\bf c$. With the randomly initialized weight (as shown in Figure \ref{fig:schema}$\bf b$), the data points are more evenly distributed, while in Figure \ref{fig:schema}$\bf c$ where the Org-Mol pre-trained weight is applied, the data points with similar molecular weights tend to cluster together, elucidating that a simple structure-property relationship has been constructed by the Org-Mol pre-trained model. As detailed below, to accurately predict physical properties of various organic compounds, we carry out the fine-tuning process for each physical property of interest so that a more specific and sophisticated structure-property relationship can be established.  

\subsection{Prediction of the electrical property}
\begin{figure}[h]
  \centering
  \includegraphics[width=16.5cm]{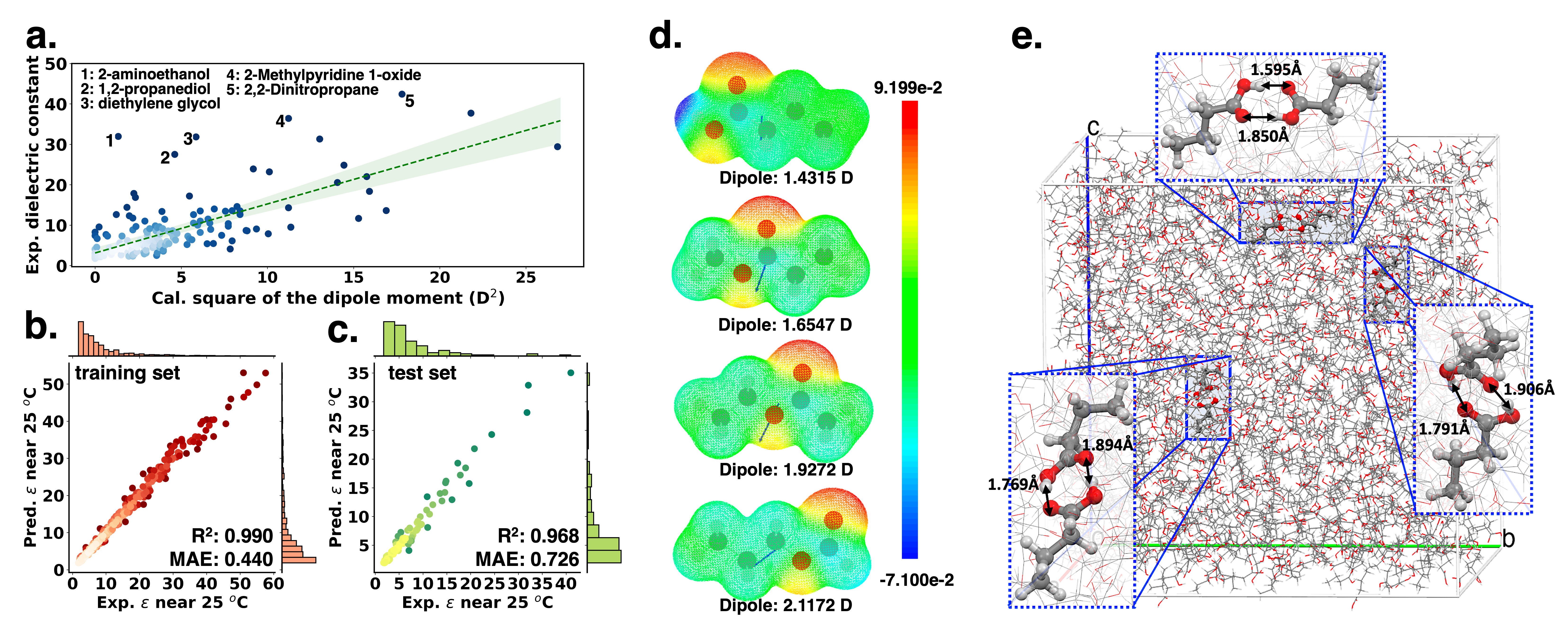}
  \caption{$\bf a.$ The experimental dielectric constant with respect the the square of the calculated molecular dipole moment of randomly picked systems. Dashed green line denotes the linear fit and shaded area denotes three standard deviations. $\bf b, c$ The correlation between Org-Mol predicted dielectric constants and experimental counterparts for the training set and the test set, respectively. $\bf d.$ Electrostatic charge distribution and molecular dipole moment of butanoic acid, methyl propanoate, ethyl acetate, and propyl methanoate (from top to bottom). $\bf e.$ Snapshots of the MD simulation of butanoic acid that indicates the existence of the dimer structure.}
  \label{fig:eps_tot}
\end{figure}

The dielectric constant is chosen as a representative to validate the performance of Org-Mol on predicting the electrical property, which plays a crucial role in the field of energy materials, especially in the areas of electrochemical energy storage as well as the development of novel immersion coolants.\cite{zhang2025immersion} The static dielectric constant $\varepsilon$ of a given system is the value of the frequency-dependent
dielectric function in the limit of zero frequency.\cite{maji2021dielectric} The numerical value
of $\varepsilon_{\textrm S}$ can be evaluated via the fluctuations of the total dipole moment of the system, which equals the summation of the molecular dipole
moments (${\mathbf\mu}_i$), i.e., ${\bf M} = \sum_i {\mathbf\mu}_i$. Mathematically, one has\cite{neumann1983dipole,neumann1984consistent}
\begin{eqnarray}
\varepsilon = 1 + \frac{\langle{\bf M}^2\rangle - \langle{\bf M}\rangle^2}{3\epsilon_0 V k_{\textrm B} T}
\end{eqnarray}
where $\epsilon_0$ is vacuum permittivity, $V$ the volume, $k_{\textrm B}$ the Boltzmann constant, and $T$ the temperature. For systems with a totally random and uncorrelated arrangement of dipoles, $ \langle{\bf M}\rangle^2=0$ and $\varepsilon$ are proportional to the square of the average molecular dipole moment. However, such straightforward relationship between the static dielectric constant of a bulk organic system and the molecular dipole moment can merely be satisfied in practice due to the unpredictable orientation of the single molecular dipoles. As shown in Figure \ref{fig:eps_tot}$\bf a$, the correlation coefficient ($R$) between the square of the computed single molecular dipole moment (without averaging over different configurations) and the experimental dielectric constant is only 0.728. A few conspicuous outliers are those with hydroxyl group and/or amino group (as labeled in Figure \ref{fig:eps_tot}$\bf a$), which introduce the formation of the hydrogen bond and the corresponding oriented arrangement of the dipoles. Consequently, the single molecular dipole moment, despite being straightforward to calculate, is not an appropriate metric especially for a quantitative evaluation of the static dielectric constant. 
It should be noted that previous studies have demonstrated the feasibility of accurately computing the dielectric constant molecular dynamics simulations for organic systems via a delicate approach to extract the short-ranged Kirkwood $g$-factor.\cite{zhang2016computing,yamaguchi2020simulations} Though with high accuracy, such calculations usually require remarkably long simulation time for a single system (especially for those large flexible organics), and is merely affordable to be applied in high throughput screening process.

While the static dielectric constant is a bulk property, its value for various organic compounds can be accurately predicted via the Org-Mol fine-tuned model with only the single molecular coordinates as the input. As shown in Figure \ref{fig:eps_tot}$\bf b$, for the static dielectric constant at the room temperature, the $R^2$ value of the training set (over 850 organic liquids) is 0.990, while the $R^2$ value of the test set is around 0.968, with an overall MAE 0.726 (as shown in Figure \ref{fig:eps_tot}$\bf c$). Such high accuracy lies in the fact that tremendous molecular structures are fed to the pre-trained model, and the corresponding structural characteristics are efficiently captured by the final 512 descriptors via a 15-layer transformer-based neural network. Therefore, the Org-Mol fine-tuned model can be applied for an effective screening tool to filter out potential candidates for immersion coolants with desired low dielectric constant.  

During the prediction process, it is found that the dielectric constant of a given carboxylic acid is lower than that of its isomeric esters even though the carboxilic group seems ``more polar'' as compared to the ester group. As shown in Table S1, the predicted dielectric constants for butanoic acid, methyl propanoate, ethyl acetate, and propyl methanoate are 2.990, 6.017, 6.267, and 6.973, respectively, which are highly consistent with the corresponding experimental values measured at 303K. Similar observations can be found in other carboxylic acids and their corresponding isomeric esters as demonstrated in Table S1. This seemingly counterintuitive phenomenon can be ascribed to two aspects. First, the molecular dipole moment of the carboxylic acid is actually lower than that of its isomeric esters. As shown in Figure \ref{fig:eps_tot}$\bf d$, although the carboxylic group itself possesses a more polarized charge distribution than the ester group does, the interaction between the positively charged protonic hydrogen and the negatively charged carbonyl oxygen within the carboxylic acid effectively counterbalances the overall molecular dipole moment. Second, due to the intermolecular hydrogen bond, the condensed phase carboxylic acid contains plentiful ``symmetric'' dimer structures (with remarkably low dipole moment $\sim$0.6 Debye based on DFT calculation) as shown in Figure \ref{fig:eps_tot}$\bf e$, which lowers the overall polarity and hence the dielectric constant. 

\subsection{Prediction of the mechanical and thermal properties}
We now examine the capability of Org-Mol in predicting the viscosity, one of the crucial metrics in fluid mechanics. Note that compared to the dielectric constant, the viscosity is more sensitive to the temperature. Therefore, we carry out two separate fine-tuning jobs based on the pre-trained model with two set of kinematic viscosities of various organic liquids measured at 40 \textcelsius\ and 100 \textcelsius, respectively. Note also that the relationship between the dynamic viscosity $\mu$ and the kinematic viscosity $\nu$ is $\mu = \nu\times\rho$, where $\rho$ is the density.\cite{alam2018density} To enable the prediction for both the dynamic viscosity and the kinematic viscosity as well as the transformation between them, we also carry out a multi-regression fine-tuning task to predict the density of organic compounds at 25 \textcelsius, 40 \textcelsius, 60 \textcelsius, and 100 \textcelsius.

\begin{figure}[h]
  \centering
  \includegraphics[width=8cm]{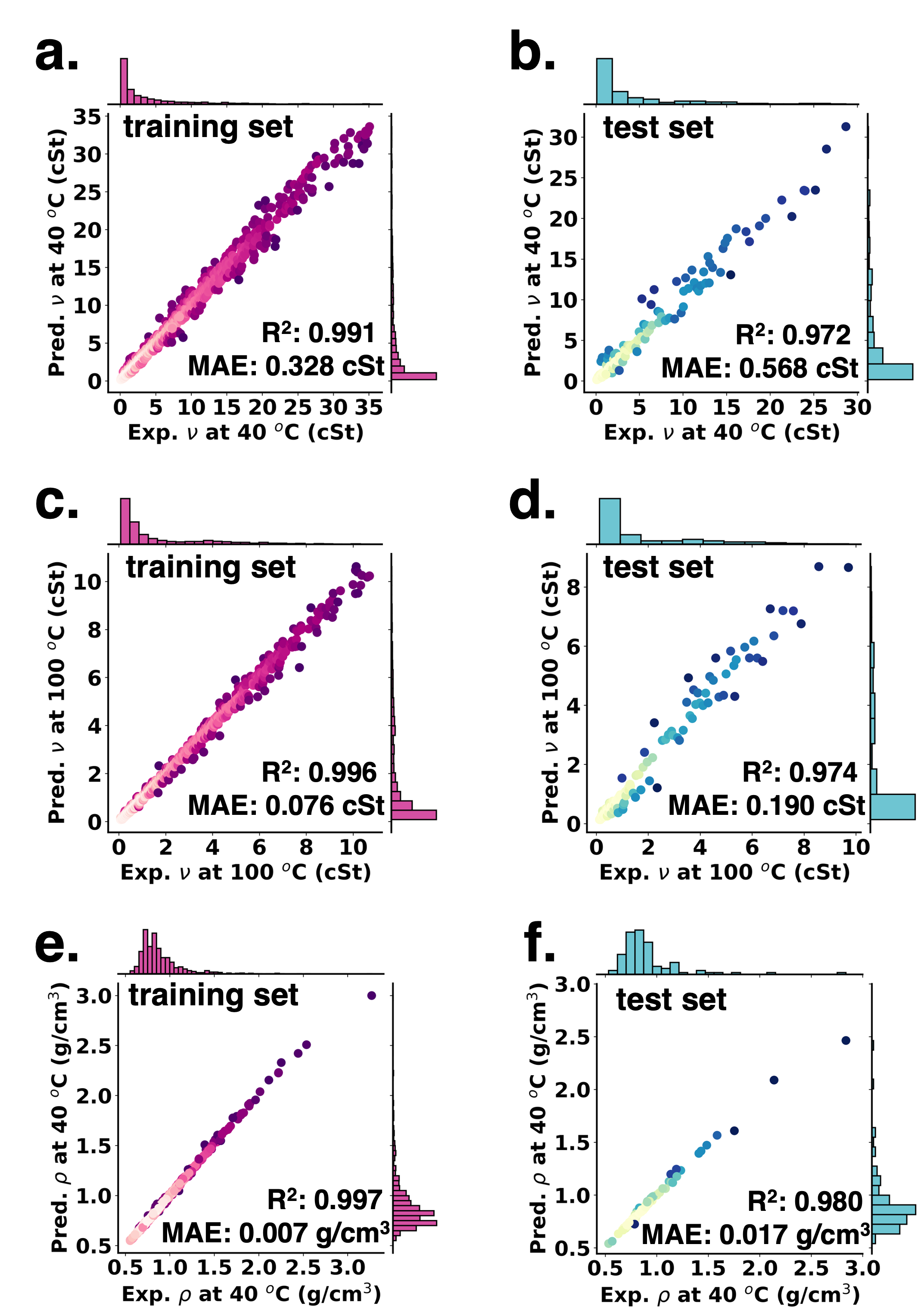}
  \caption{The correlation between Org-Mol predicted kinematic viscosity at 40 \textcelsius\ and experimental counterparts for $\bf a$ the training set and $\bf b$ the test set. The correlation between Org-Mol predicted kinematic viscosity at 100 \textcelsius\ and experimental counterparts for $\bf c$ the training set and $\bf d$ the test set. The correlation between Org-Mol predicted density at 40 \textcelsius\ and experimental counterparts for $\bf e$ the training set and $\bf f$ the test set.}
  \label{fig:visco}
\end{figure}

It can be seen in Figure \ref{fig:visco}$\bf a$-$\bf d$ that the predicted values of the kinematic viscosity at two tested temperatures are highly consistent with the corresponding experimental counterparts. The $R^2$ values for the train and test set of the kinematic viscosity at 40 \textcelsius\ are 0.991 and 0.972, respectively, while those at 100 \textcelsius\ are even higher (0.996 and 0.974, respectively). Compared to the prediction of dielectric constant and the viscosity, predicting the density is less difficult, due to the fact that the latter is a pure structural property and is less relevant to the dynamical behavior of the system. As shown in Figure \ref{fig:visco}$\bf e$-$\bf f$ and Figure S2, the predicted densities are in extremely good agreement for both the train set and the test set. The slight underestimation of densities in high-density regions can be attributed to the sparsity of experimental data in those areas. In practice, very few common organic liquids reach such elevated densities, thus mitigating any significant impact on potential material design and selection processes.

\begin{figure}[h]
  \centering
  \includegraphics[width=8cm]{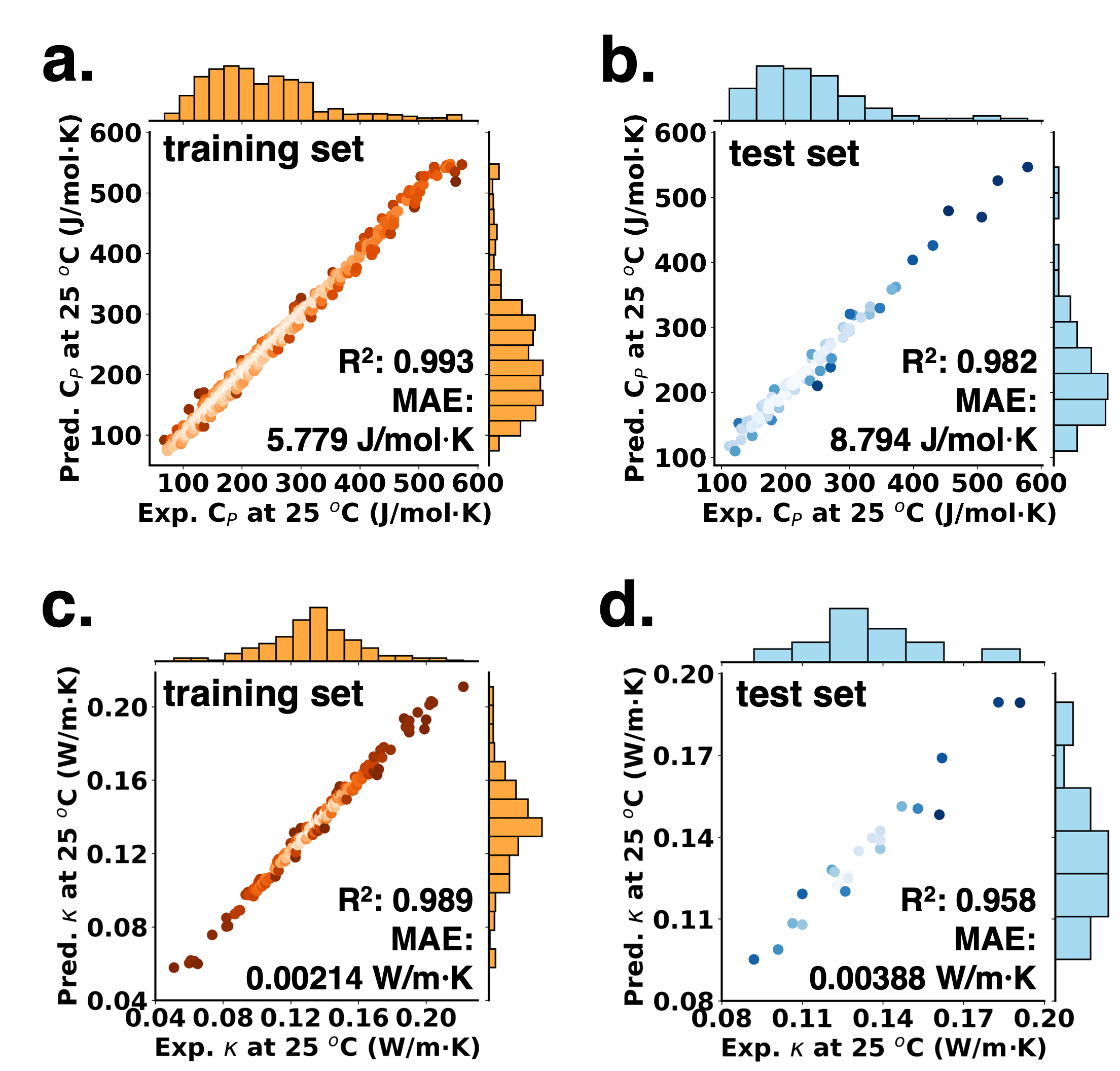}
  \caption{The correlation between Org-Mol predicted heat capacity and experimental counterparts for $\bf a$ the training set and $\bf b$ the test set. The correlation between Org-Mol predicted thermal conductivity and experimental counterparts for $\bf c$ the training set and $\bf d$ the test set.}
  \label{fig:heat_thermo}
\end{figure}

For thermal properties, we take the thermal conductivity $\kappa$ and the specific heat at constant pressure $C_p$ as two representatives to examine the performance of Org-Mol. As shown in Figure  \ref{fig:heat_thermo}, the correlation between the Org-Mol predicted values and the experimental counterparts for both these two thermal properties remains high. The $R^2$ values for the training and test sets are 0.989 and 0.958 for the heat capacity, respectively, with very small MAEs, evincing the capability of Org-Mol in the accurate prediction of thermal properties. Note that even though only 248 data points collected for thermal conductivity, the performance of Org-Mol on the test set is still satisfactory (with the the $R^2$ values 0.958 for the test set), demonstrating its strength of tackling data with relatively small sample size. 

\subsection{Comparison with reference models}
For all the aforementioned properties, we carry out the same training process via EGNN\cite{satorras2021n} and NequIP\cite{batzner20223} algorithms (with correspondingly optimized hyper parameters). The metrics for the test set are summarized in Table \ref{tbl:metric}. It can be seen that Org-Mol outperforms two reference models for all properties of interest, especially for the dielectric constant and the thermal conductivity, of which the complexity and the sparse experimental data raise challenges for the training process. Significant improvement of all metrics are observed via Org-Mol for these two properties as compared to EGNN and NequIP, highlighting the strength of our pre-trained model in tackling the complex and/or sparsed properties . 

\begin{table}
\caption{Metrics of Org-Mol, EGNN, and NequIP for the dielectric constant $\varepsilon$, the kinematic viscosity $\nu$ at 40  \textcelsius\ and 100 \textcelsius, the heat capacity $C_p$, and the thermal conductivity $\kappa$.\textsuperscript{\emph{a}} }
\label{tbl:metric}
\begin{tabular}{c|c|c|c|c|c|c}
\hline\hline
 Metric & Model & $\varepsilon$ & $\nu$ at 40 \textcelsius\ & $\nu$ at 100 \textcelsius\  & $C_p$ & $\kappa$ \\
    \hline
  & our work & 0.968 &	0.972	& 0.974 & 0.982 & 0.958\\
  $R^2$ & EGNN  & 0.869	 & 0.956	&	0.945	&0.981	&0.918 \\
 & NequIP & 0.885	&	0.948	&	0.936	& 0.980	 & 0.885\\
    \hline
  & our work & 0.726 & 0.568	&	0.190	& 8.794	&3.878E-03\\
  MAE\textsuperscript{\emph{b}} & EGNN  & 1.111	&	0.615	&0.245	&9.190	& 5.040E-03 \\
 & NequIP     & 1.153 & 0.668    &0.263	 &9.366	 & 5.420E-03\\
    \hline
  & our work &1.220 &0.999 &	0.336 & 12.073	&4.921E-03\\
  RMSE\textsuperscript{\emph{b}} & EGNN  & 2.470	&	1.261	& 0.491	&12.286	& 6.529E-03 \\
 & NequIP     & 2.280 &1.362   & 0.522	 &12.420	 & 7.579E-03\\
    \hline
  & our work & 9.233 & 22.467	&	12.528	& 3.978	& 2.911\\
  MAPE(\%) & EGNN  & 13.165	&	23.482	& 14.448	&4.172	& 3.704\\
 & NequIP     & 15.132 & 24.732  &14.894	 & 4.348 & 3.984\\
    \hline\hline

  \end{tabular}\\
\textsuperscript{\emph{a}} Tested properties correspond to the value measured at room temperatue unless otherwise specified.;
\textsuperscript{\emph{b}} Units for $\nu$, $C_p$, and $\kappa$ are cSt, J/mol$\cdot$K, and W/m$\cdot$K, respectively, while those for $\varepsilon$ is unitless.
\end{table}

The heat capacity of organic compounds, being a fairly predictable characteristic, has been thoroughly investigated using a multitude of ML techniques. As summarized in Table \ref{tbl:heatcap}, while previously reported models have gain relatively satisfying accuracies in predicting the heat capacity of various organics, Org-Mol achieves the highest accuracy among all applied algorithms (with marginal improvement as compared to EGNN and NequIP) over a larger and more diverse dataset. 

\begin{table}
\caption{Comparison of previously reported models and this work for the heat capacity $C_p$.}
\label{tbl:heatcap}
\begin{tabular}{c|c|cc|cc|cc|cc|c }
\hline\hline
No. of  & $T$  & \multicolumn{2}{c|}{$R^2$}  &  \multicolumn{2}{c|}{MAE\textsuperscript{\emph{a}} }  & \multicolumn{2}{c|}{RMSE\textsuperscript{\emph{a}} } & \multicolumn{2}{c|}{MAPE (\%)} &  \\ \cline{3-10}
compounds & range (K) & Train & Test & Train & Test & Train & Test & Train & Test & Ref\\ \hline
1009 & 298.15 & 0.994 & 0.982 & 5.779 & 8.794 & 7.965 &12.073 & 2.806 & 3.978 & this work \\
855 & 250-335 & 0.991 & 0.974 & 4.395 & 9.385 & 7.275 &12.322 & 2.184 & 5.248 & GBRT\cite{shan2024develop} \\
706 & 298.15 & 0.965 & 0.972 & - & - &16.137 &16.632 & 4.77 &5.22 &ANFIS\cite{khajeh2012quantitative} \\
\hline\hline
  \end{tabular}\\
\textsuperscript{\emph{a}} Values are shown in the unit of J/mol$\cdot$K.
\end{table}

\subsection{Experimental validation for the discovery of immersion cooling candidates}
\begin{figure}[h]
  \centering
  \includegraphics[width=16.5cm]{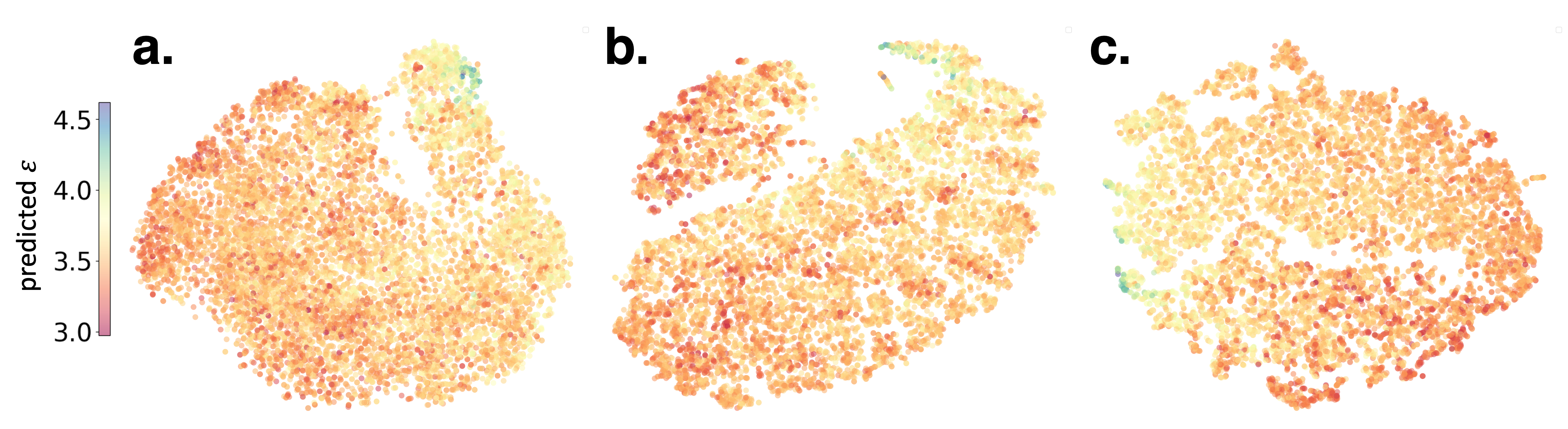}
  \caption{Distributions of the dielectric constant of monoesters in 2D projections of Org-Mol embedding layer via t-SNE with $\bf a$ randomly initialized weight, $\bf b$ the pre-trained weight, and $\bf c$ the fine-tuned weight.}
  \label{fig:eps-tsne}
\end{figure}

As a proof-of-concept of how these property-prediction models can accelerate the discovery of new materials as well as to experimentally validate their accuracy, we apply the Org-Mol fined-tuned models to systematically screen out promising candidates for immersion coolants, which require low dielectric constant (ideally lower than 3.20 at room temperature), low kinematic viscosity (preferably lower than 12 cSt at 40 \textcelsius), high thermal conductivity (preferably higher than 0.140 W/m$\cdot$K), and environmental friendliness. We focus on synthetic esters for the design of novel immersion coolants due to its extraordinary biodegradability. One of the plausible problems for synthetic esters is that the involvement of the ester group introduces the polarity and thus increase the dielectric constant. Therefore, we apply the dielectric constant as a primary criterion and screen out over 6 million saturated monoesters and dibasic esters (constructed from saturated acid and alcohol checked out from PubChem) with the number of carbon atoms lower or equal to 30. Interestingly, all tested dibasic esters possess dielectric constants larger than 3.50, while 8569 monoesters meet the criterion and the corresponding t-SNE analysis is applied to the predicted dielectric constants of these monoesters to further justify the effectiveness of both the pre-trained and the fine-tuned model. As shown in Figure \ref{fig:eps-tsne}$\bf a$, with the randomly initialized weight, data points are evenly distributed with the 2D projected space. The distribution after pre-training, as seen in Figure  \ref{fig:eps-tsne}$\bf b$, is capable of effectively distinguishing certain structures. For instance, in the upper-right corner of Figure \ref{fig:eps-tsne}$\bf b$, molecules with higher dielectric constants are grouped together. Upon fine-tuning, a noticeable decreasing trend of the dielectric constant values emerges from left to right in Figure \ref{fig:eps-tsne}$\bf c$ as compared to the pre-trained model, which indicates that the combination of the pre-training and fine-tuning process can more accurately differentiate molecules based on their dielectric constants. 

Monoesters with dielectric constants lower than 3.20 are then fed to the kinetic viscosity model and then the thermal conductivity model. Based on the aforementioned criteria of the kinematic viscosity and thermal conductivity, 461 monoesters are finally filtered out. By further taking into the consideration of the synthetic accessibility, we synthesize two potential candidates and their physical properties are experimentally measured (of which the experimental characterization is provided in the Supporting Information). The whole screening process is summarized in Scheme \ref{scm:ester}. As shown in Table \ref{tbl:exp_val}, the predicted values for all tested properties of these two candidates are in good agreement with the experimental results. A slight systematic overestimation is observed for both the kinematic viscosity and the thermal conductivity, presumably due to the inconsistency of the measuring methods between the current work and the referenced works contributing to the data set. (See Supporting Information for testing methods applied in this work.) It should be noticed that even though the thermal conductivity of the second candidates is slightly lower than the preferred value (0.140 W/m$\cdot$K), its low dielectric constant ensures superior electrical insulation, which is particularly crucial for practical applications. Overall, the Org-Mol fine-tuned models succeed in screening out promising candidates for immersion coolants over millions of candidates, and can be further applied in the discovery of other energy-saving materials. 
\begin{scheme}[!h]
  \includegraphics[width=8cm]{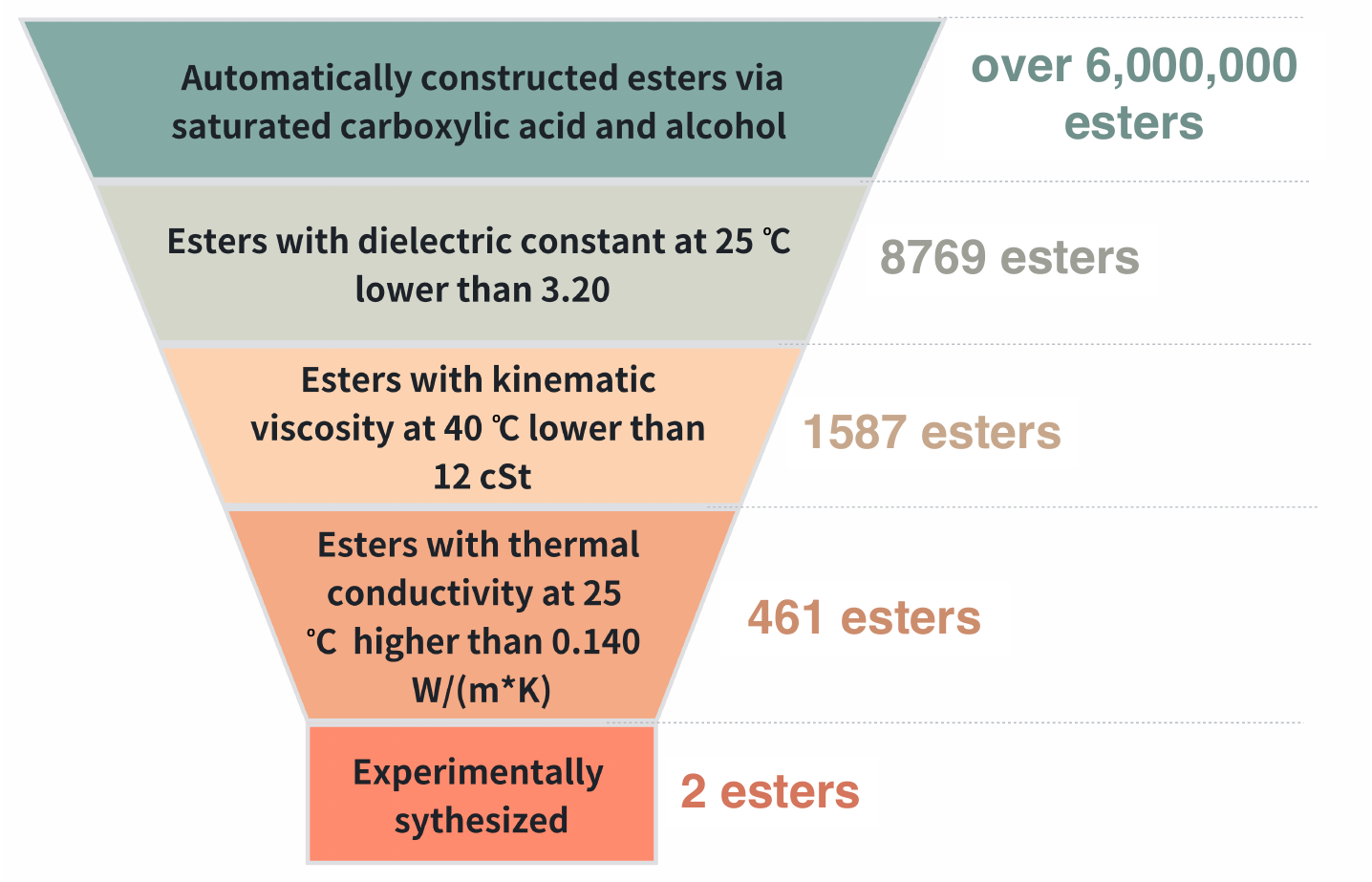}
  \caption{Schematic graph of the screening procedure.}\label{scm:ester}
\end{scheme}

\begin{table}
\caption{Predicted and experimentally measured dielectric constants $\varepsilon$, kinematic viscosities $\nu$ at 40  \textcelsius\ and 100 \textcelsius, and thermal conductivities $\kappa$ for two filtered-out immersion cooling candidates. }
\label{tbl:exp_val}
\begin{tabular}{c|c|c|c|c|c|c|c|c}
\hline\hline
& \multicolumn{2}{c|}{$\varepsilon$}  &  \multicolumn{2}{c|}{$\nu$ at 40 \textcelsius\ (cSt)}  & \multicolumn{2}{c|}{$\nu$ at 100 \textcelsius\ (cSt)} & \multicolumn{2}{c}{$\kappa$ (W/m$\cdot$K)}  \\
   \cline{2-9}
  Candidates & Exp. & Pred. & Exp. & Pred. & Exp. & Pred. & Exp. & Pred. \\ \hline
 \bf{Ester 1} & 2.80 &	3.13 &	6.41 &	7.18 &	1.98 &	2.23 &	0.141	 & 0.144 \\
  \bf{Ester 2} & 2.70 &	2.87 &	9.67 &	10.84 &	2.59 & 2.98	& 0.138 & 0.141 \\ \hline\hline
  \end{tabular}\\
\end{table}

\section{Conclusion}
In conclusion, we have applied the framework of the molecular representation learning and developed the Org-Mol pre-trained model, which is specifically designed for organic compounds on a vast dataset of 60 million semi-empirically optimized structures. Leveraging the power of 3D transformer-based algorithms, we have successfully fine-tuned the Org-Mol pre-trained model , achieving $R^2$ values exceeding 0.95 for test sets across various physical properties of organic compounds. This high level of accuracy not only surpasses current state-of-the-art models but also offers a practical solution to the challenges posed by traditional experimental methods, which are often costly and time-consuming.

The application of Org-Mol in high-throughput screening has demonstrated its capability to enable the rapid identification of potential immersion coolants from a pool of millions of automatically constructed ester molecules. The experimental validation of two promising candidates further underscores the practical utility of our approach in accelerating the discovery of new materials for energy-efficient applications.

In summary, the Org-Mol model represents a powerful tool in the predictive modeling of organic compounds, with broad implications for the development of energy-saving materials. Its success in predicting electrical, mechanical, and thermal properties paves the way for the design of novel amorphous or liquid phase organic materials with tailored properties, contributing to the advancement of sustainable energy solutions and environmental compatibility.

\begin{acknowledgement}
This work was supported by research grants from China Petroleum \& Chemical Corp (funding number 124014).
\end{acknowledgement}

\begin{suppinfo}

Detailed fine-tuning processes and hyper-parameters for each physical property tested in the work. Experimental dielectric constants of some carboxylic acids and their isomeric esters. Correlation between the predicted and experimental density at 25 \textcelsius, 60 \textcelsius, and 100 \textcelsius. Experimental procedures, characterization data, and testing methods for two synthesized esters. 

\end{suppinfo}

\bibliography{achemso-demo}

\end{document}


\newpage
\section{Org-Mol fine-tuning details}
\begin{figure}
  \centering
  \includegraphics[width=8cm]{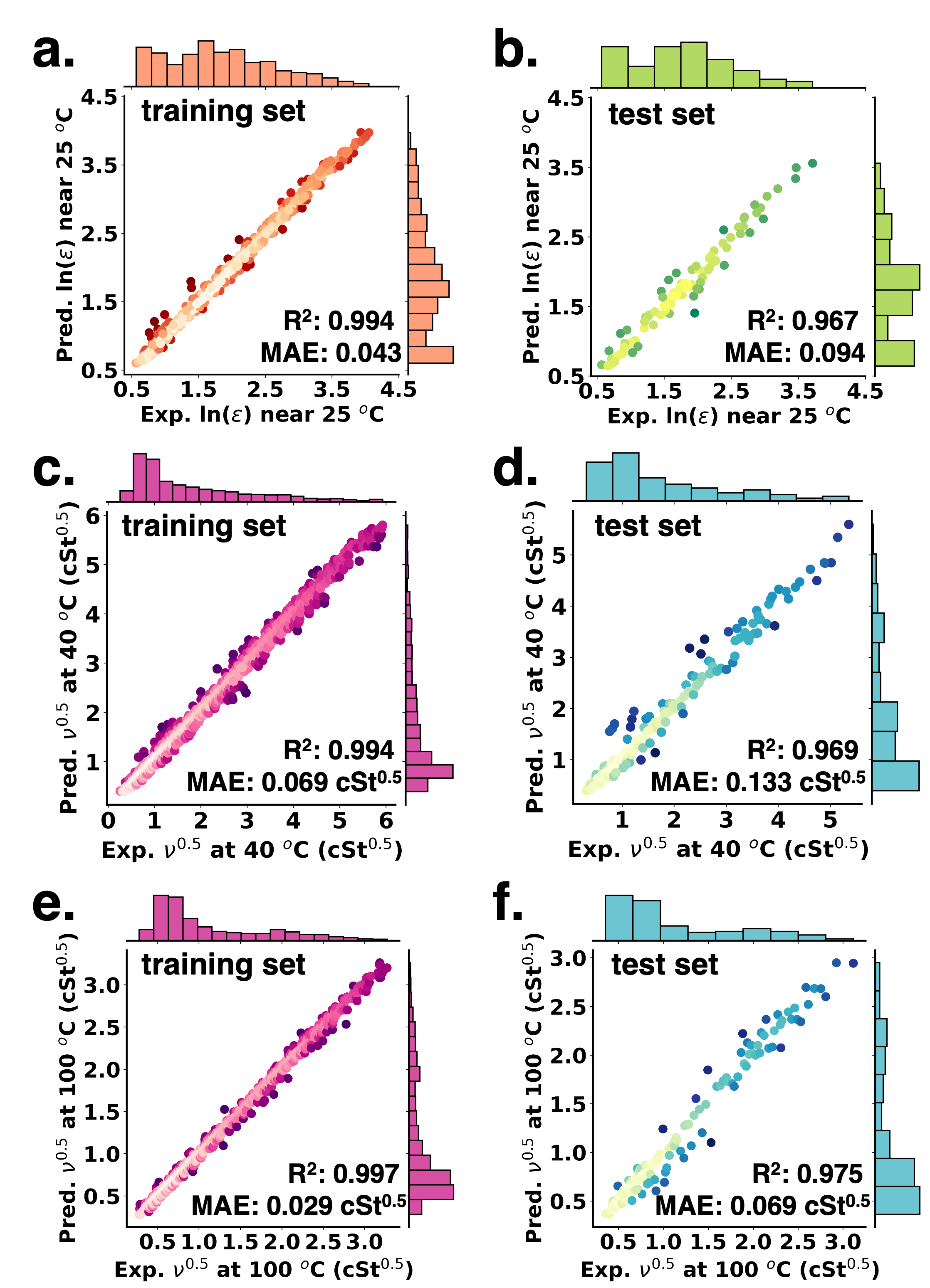}
  \caption{The correlation between Org-Mol the predicted logarithm of the dielectric constant near 25 \textcelsius\ and experimental counterparts for $\bf c$ the training set and $\bf d$ the test set. The correlation between Org-Mol the predicted square root of the kinematic viscosity at 40 \textcelsius\ and experimental counterparts for $\bf c$ the training set and $\bf d$ the test set. the predicted square root of the kinematic at 100 \textcelsius\ and experimental counterparts for $\bf e$ the training set and $\bf f$ the test set.}
  \label{fig:raw}
\end{figure}

The Org-Mol pre-trained model is fine-tuned via four experimentally measured properties, namely, the dielectric constant near 25 \textcelsius, kinematic viscosities at 40 \textcelsius\ and 100 \textcelsius, densities at 25 \textcelsius, 40 \textcelsius,  60 \textcelsius, and 100 \textcelsius\ (as a multi-regression task), the heat capacity at 25 \textcelsius, and the thermal conductivity at 25 \textcelsius. The pre-processing of the training data is applied for the dielectric constant and the kinematic viscosities due to the unbalanced distribution of these properties, i.e., data points are clustered in the low value range. For the dielectric constant, we take the natural logarithm of the original data for training. For the kinematic viscosities at both temperatures, we take the square root of the original data for training. The raw training results for these properties are depicted in Figure \ref{fig:raw}. Note that Org-Mol automatically cleans up anomal data with 3$\sigma$ threshold in the training set and those data are excluded in all training sets statistics. The same pre-processing as mentioned above is applied when training the EGNN model and the NequIP model so as to make a fair comparison. The pre-trained Org-Mol model is accessible via {\fontfamily{qcr}\selectfont https://funmg.dp.tech/orgmol}, while the original training and test data sets of all aforementioned properties (including the corresponding PM6-optimized 3D coordinates) are compressed in the supplementary file {\fontfamily{qcr}\selectfont data.zip} with a {\fontfamily{qcr}\selectfont pickle} format.   

\begin{figure}
  \centering
  \includegraphics[width=8cm]{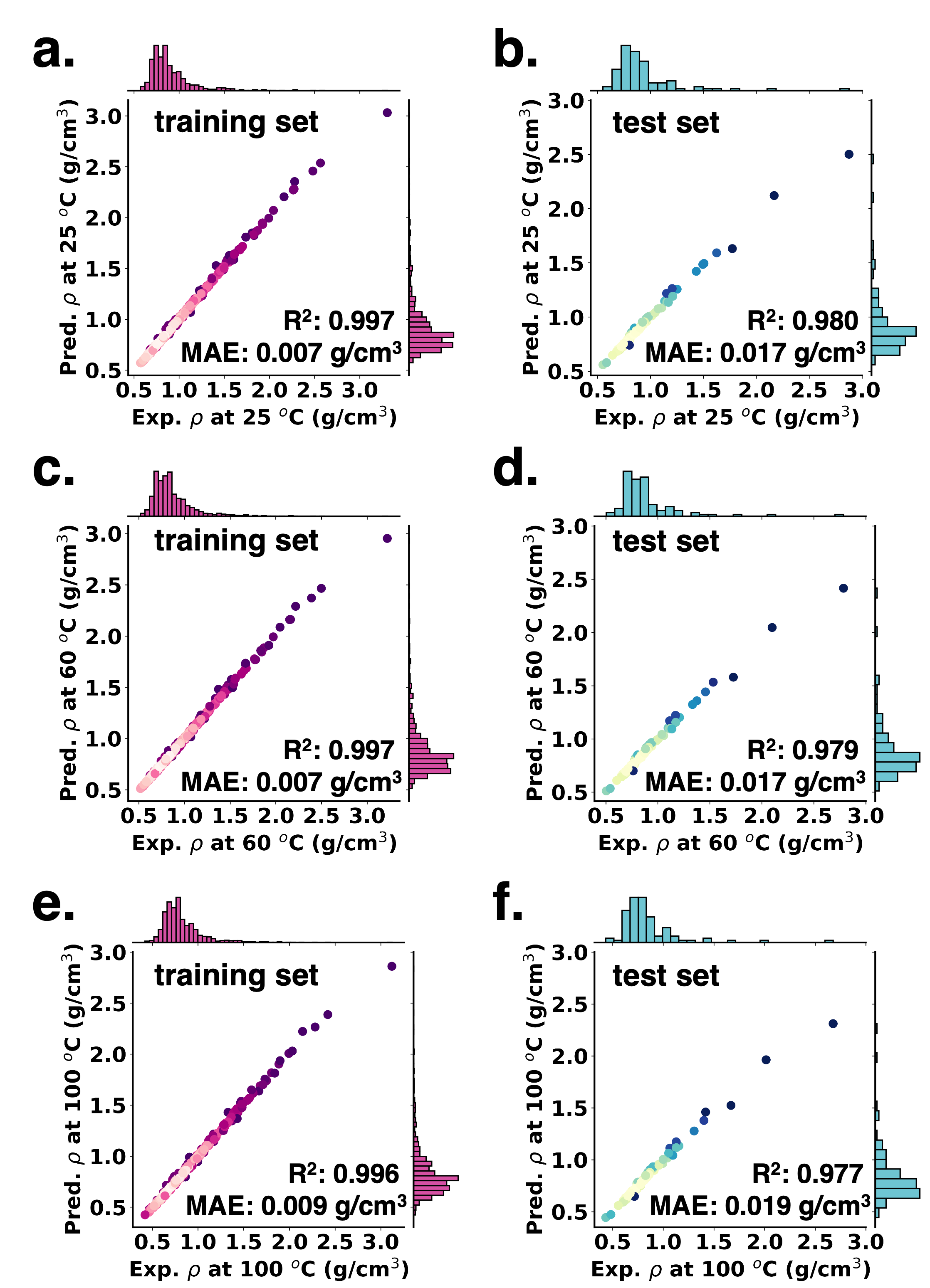}
  \caption{The correlation between Org-Mol predicted density at 25 \textcelsius\ and experimental counterparts for $\bf a$ the training set and $\bf b$ the test set. The correlation between Org-Mol predicted density at 60 \textcelsius\ and experimental counterparts for $\bf c$ the training set and $\bf d$ the test set. The correlation between Org-Mol predicted density at 100 \textcelsius\ and experimental counterparts for $\bf e$ the training set and $\bf f$ the test set.}
  \label{fig:density}
\end{figure}

\newpage
The random seed for all fine-tuning processes is 42. Hyper-parameters for each fine-tuning process is given as follows:
\begin{enumerate}
    \item Dielectric constant near 25 \textcelsius
\begin{spverbatim}
amp: true
anomaly_clean: true
batch_size: 32
early_stopping: 30
epochs: 150
learning_rate: 0.00029
logger_level: 1
max_epochs: 100
max_norm: 5.0
metrics: r2
num_classes: 1
patience: 40
remove_hs: false
seed: 42
smi_strict: false
split: random_5fold
split_method: 5fold_random
split_seed: 42
target_normalize: auto
task: regression
warmup_ratio: 0.06
\end{spverbatim}

\item Kinematic viscosity at 40 \textcelsius
\begin{spverbatim}
amp: true
anomaly_clean: true
batch_size: 32
early_stopping: 30
epochs: 150
learning_rate: 0.00028
logger_level: 1
max_epochs: 100
max_norm: 5.0
metrics: r2
num_classes: 1
patience: 40
remove_hs: false
seed: 42
smi_strict: false
split: random_5fold
split_method: 5fold_random
split_seed: 42
target_normalize: auto
task: regression
warmup_ratio: 0.06
\end{spverbatim}

\item Kinematic viscosity at 100 \textcelsius
\begin{spverbatim}
amp: true
anomaly_clean: true
batch_size: 32
early_stopping: 30
epochs: 150
learning_rate: 0.00030
logger_level: 1
max_epochs: 100
max_norm: 5.0
metrics: r2
num_classes: 1
patience: 40
remove_hs: false
seed: 42
smi_strict: false
split: random_5fold
split_method: 5fold_random
split_seed: 42
target_normalize: auto
task: regression
warmup_ratio: 0.06
\end{spverbatim}

\item Density at 25 \textcelsius, 40 \textcelsius, 60 \textcelsius, and 100 \textcelsius
\begin{spverbatim}
amp: true
anomaly_clean: true
batch_size: 32
early_stopping: 30
epochs: 150
learning_rate: 0.00034
logger_level: 1
max_epochs: 100
max_norm: 5.0
metrics: mae,mse,r2
num_classes: 4
patience: 40
remove_hs: false
seed: 42
smi_strict: false
split: random_5fold
split_method: 5fold_random
split_seed: 42
target_normalize: auto
task: multilabel_regression
warmup_ratio: 0.06
\end{spverbatim}

\item Heat capacity at 25 \textcelsius
\begin{spverbatim}
amp: true
anomaly_clean: true
batch_size: 32
early_stopping: 30
epochs: 150
learning_rate: 0.000275
logger_level: 1
max_epochs: 100
max_norm: 5.0
metrics: r2
num_classes: 1
patience: 40
remove_hs: false
seed: 42
smi_strict: false
split: random_5fold
split_method: 5fold_random
split_seed: 42
target_normalize: auto
task: regression
warmup_ratio: 0.06
\end{spverbatim}

\item Thermal conductivity at 25 \textcelsius
\begin{spverbatim}
amp: true
anomaly_clean: true
batch_size: 32
early_stopping: 30
epochs: 130
learning_rate: 0.00032
logger_level: 1
max_epochs: 100
max_norm: 5.0
metrics: mae,r2
num_classes: 1
patience: 40
remove_hs: false
seed: 42
smi_strict: false
split: random_5fold
split_method: 5fold_random
split_seed: 42
target_normalize: auto
task: regression
warmup_ratio: 0.06
\end{spverbatim}
\end{enumerate}

\section{Experimental and predicted dielectric constants for carboxylic acids and their isomeric esters}
We compare the dielectric constants of two carboxylic acids and those of the corresponding isomeric esters in Table \ref{tbl:eps}. All experimental values listed in Table \ref{tbl:eps} are taken from Ref.~\citenum{LandoltBornstein1991:sm_lbs_978-3-540-47619-1_2} and measured at 303 K. Values measured at 298 K for the listed compounds are incomplete and hence we take the values measured at 303 K to make a fair comparison. The dielectric constants of common organic liquids only change slightly withn 5$\sim$10 K temerature range and therefore the experimental values measured at 303 K are comparable to those predicted near 298 K, i.e., 25 \textcelsius. The 3D coordinates optimized via PM6 method of these compounds are zipped in the supplementary coordinates file.

\begin{table}
\caption{Experimental and predicted dielectric constants for carboxylic acids and their isomeric esters.}
\label{tbl:eps}
\begin{tabular}{cccc}
\hline\hline
Compound & SMILES & Exp. $\varepsilon$ & Pred. $\varepsilon$ \\
    \hline
  Butyric acid      & CCCC(=O)O     & 2.8646 & 2.990 \\
  Methyl propionate & CCC(=O)OC     & 5.942  & 6.017 \\
  Ethyl acetate     & CC(=O)OCC     & 6.065  & 6.266 \\
  Propyl formate    & C(=O)OCCC     & 6.92   & 6.973 \\
  Octanoic acid     & CCCCCCCC(=O)O & 2.5181 & 2.498 \\
  Ethyl hexanoate   & CCCCCC(=O)OCC & 4.57   & 4.581 \\
  Hexyl acetate     & CC(=O)OCCCCCC & 4.43   & 4.171 \\
  \hline\hline

  \end{tabular}\\
\end{table}

\section{Experimental characterization and property measurements of two immersion cooling candidates}
\subsection{Experimental characterization}
{\bf{Ester 1:}} 1H-NMR (700 MHz, 
CDCl3) $\delta$ (ppm) = 4.05 (t, J = 7 Hz, 2H), 2.32-2.25 
(m, 1H), 1.63-1.51 (m, 4H), 1.49-1.35 (m, 2H), 1.30-1.15 (m, 
26H), 0.88-0.81 (m, 9H). 13C-NMR (175 MHz, CDCl3) $\delta$ 
(ppm) = 176.9, 64.3, 46.0, 32.8, 32.5, 32.1, 31.9, 29.9, 
29.8, 29.7, 29.5, 29.4, 28.9, 27.6, 26.2, 22.9, 22.8, 22.7, 
14.3, 14.2, 14.1. (The 3D coordinates optimized via PM6 method of these compounds are zipped in the supplementary coordinates file.) 

\begin{figure}
  \centering
  \includegraphics[width=16.5cm]{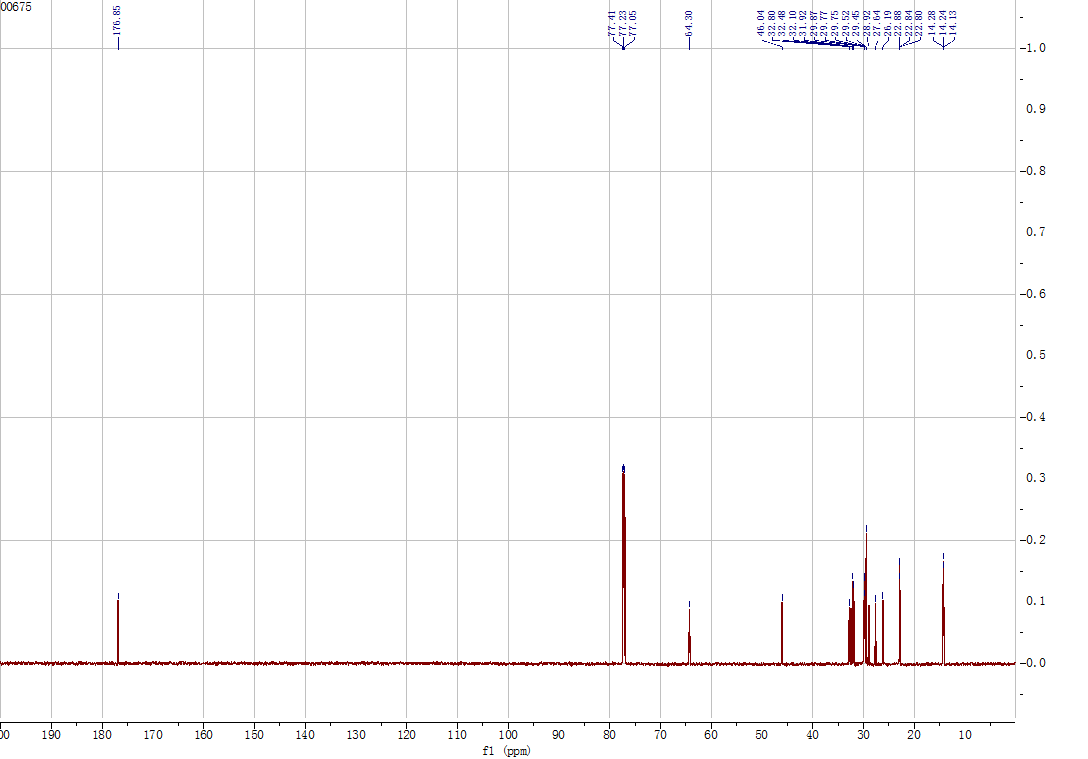}
  \caption{Carbon-13 nuclear magnetic resonance (13C-NMR) spectroscopy of {\bf{Ester 1}}.}
  \label{fig:1-C-NMR}
\end{figure}

\begin{figure}
  \centering
  \includegraphics[width=16.5cm]{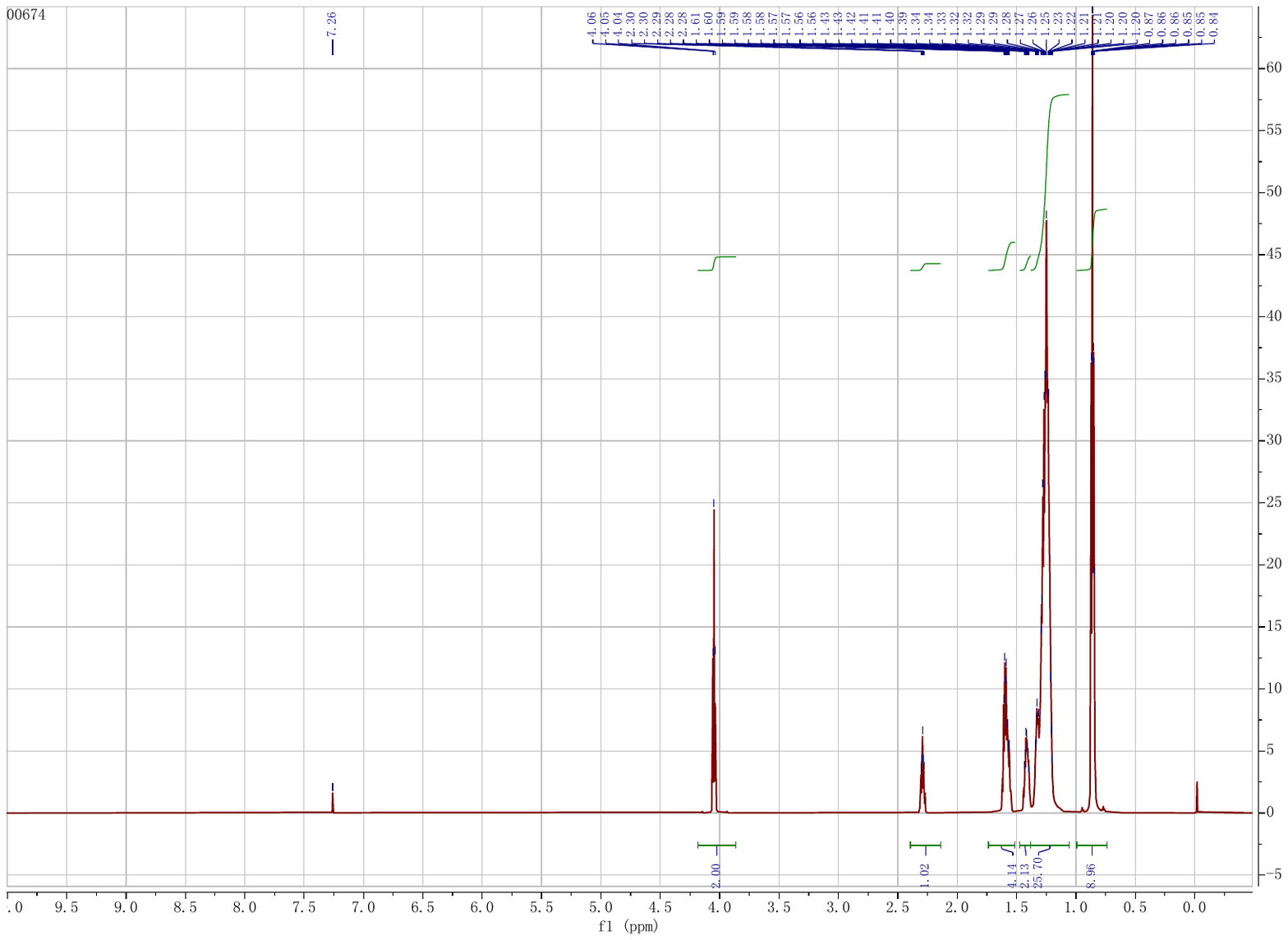}
  \caption{Proton nuclear magnetic resonance (1H-NMR) spectroscopy of {\bf{Ester 1}}.}
  \label{fig:1-H-NMR}
\end{figure}

\newpage
{\bf{Ester 2:}} 
1H-NMR 
(700 MHz, CDCl3) 
$\delta$ 
(ppm) 
= 4.14-4.05 (m, 2H), 2.33-2.26 (m, 1H), 1.70-1.46 (m, 5H), 1.45-1.36 (m, 
3H), 1.35-1.18 (m, 23H), 1.17-1.07 (m, 3H), 0.89 (d, J = 6.3 
Hz, 3H), 0.85-0.81 (m, 12H). 13C-NMR (175 MHz, CDCl3) 
$\delta$ (ppm) = 176.9, 62.7, 46.1, 39.4, 37.4, 35.9, 32.8, 
32.1, 31.9, 30.0, 29.8, 29.7, 29.5, 29.4, 28.2, 27.7, 27.6, 
24.9, 22.9, 22.8, 22.7, 19.6, 14.3, 14.2. HRMS (FT-ICRMS) 
calcd for C26H53O2 (M+H): 397.4040, found: 397.4041; calcd 
for C26H52NaO2 (M+Na): 419.3860, found: 419.3859.

\begin{figure}
  \centering
  \includegraphics[width=16.5cm]{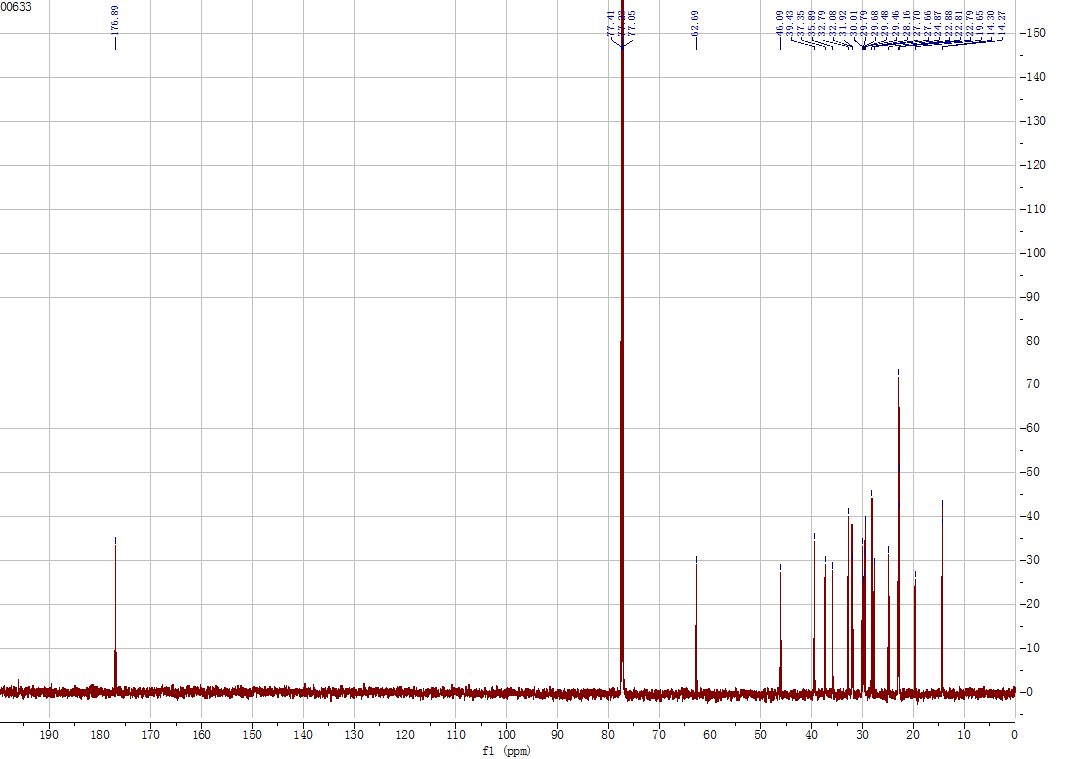}
  \caption{Carbon-13 nuclear magnetic resonance (13C-NMR) spectroscopy of {\bf{Ester 2}}.}
  \label{fig:2-C-NMR}
\end{figure}

\begin{figure}
  \centering
  \includegraphics[width=16.5cm]{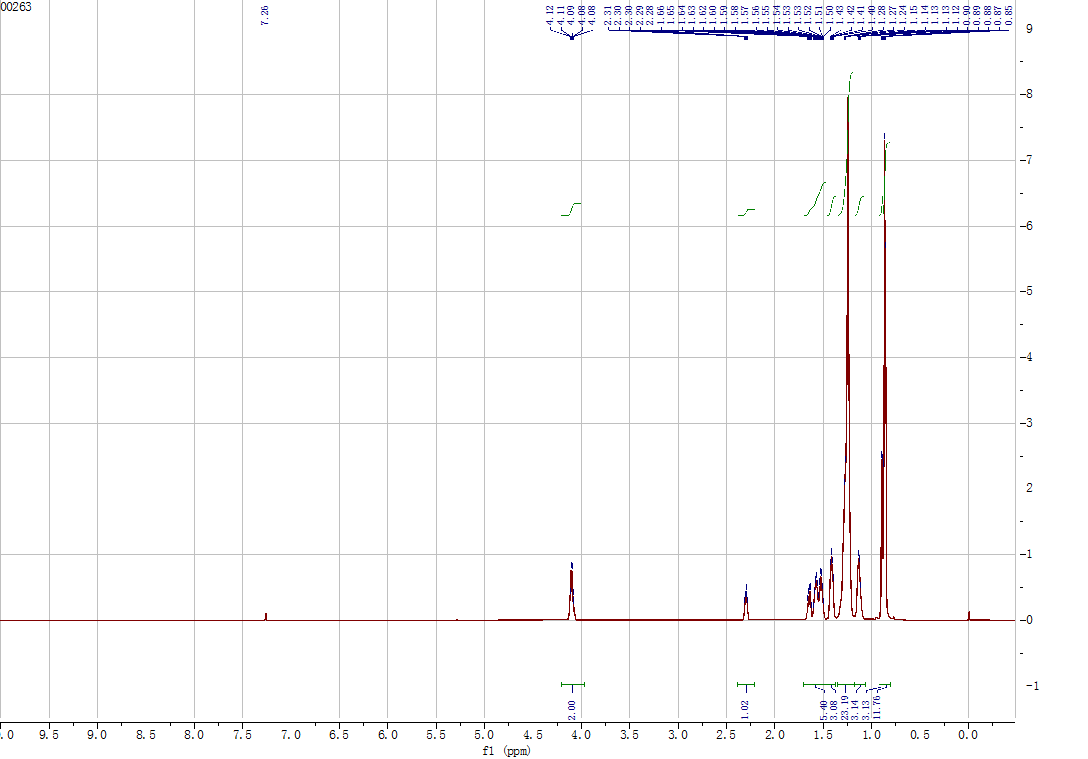}
  \caption{Proton nuclear magnetic resonance (1H-NMR) spectroscopy of {\bf{Ester 2}}.}
  \label{fig:2-H-NMR}
\end{figure}

\begin{figure}
  \centering
  \includegraphics[width=16.5cm]{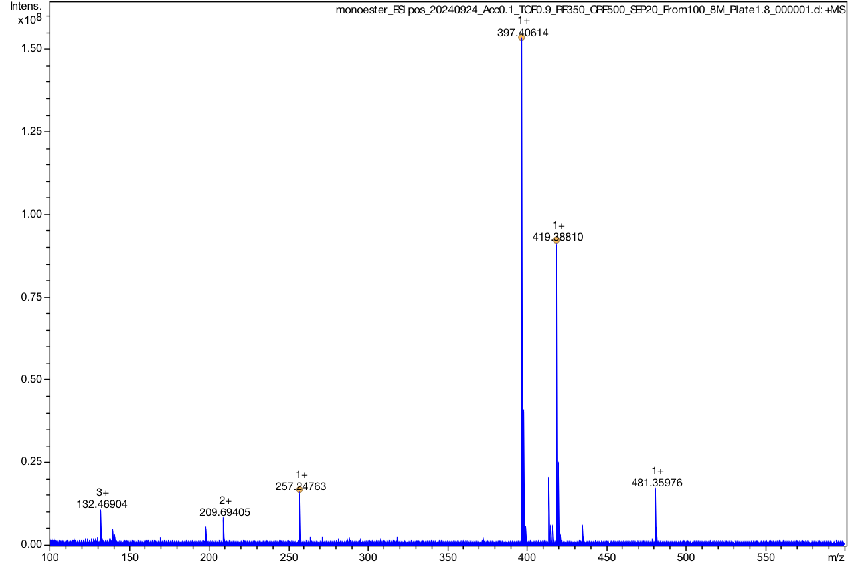}
  \caption{Electrospray ionization Fourier transform ion cyclotron resonance mass spectrometry of {\bf{Ester 2}}.}
  \label{fig:2-MS}
\end{figure}

\begin{sidewaystable*}
\caption{Testing details for dielectric constants, kinematic viscosities, and thermal conductivities of two ester compounds.}
\label{tbl:exp}
\begin{tabularx}{\textwidth}{cXXXX}
\hline\hline
Property & Test method & Device name and model number & Manufacture & Notes\\
    \hline
Dielectric constant   & IEC 60247:2004    & PS-2001A Insulating oil dielectric loss and volume resistivity tester & Baoding Push Electrical Manufacturing Co., Ltd. & Insulating liquids—measurement of relative permittivity,dielectric dissipation factor and d.c.resistivity \\ \hline
Kinematic viscosity   & ASTM D7279: testing kinematic viscosity using a Houillon viscometer     & S-flow 1250 Automated Houillon Viscometer  & Omnitek B. V. & Standard test method for kinematic viscosity of transparent and opaque liquids by automated Houillon viscometer \\ \hline
Thermal conductivity  & ASTM D7896: transient hot-wire method    & TC3100L Thermal Conductivity Meter  & XIATECH Electronic Technology Co., Ltd. & Standard test method for thermal conductivity, thermal diffusivity, and volumetric heat capacity of engine coolants and related fluids by transient hot wire liquid thermal conductivity method
\\
  \hline\hline
  \end{tabularx}%
\end{sidewaystable*}%

\newpage
\bibliography{achemso-demo}